\documentclass[useAMS,usenatbib,usegraphicx]{mn2e}
\usepackage{amsmath}
\usepackage{graphicx}

\def\simlt{\lower.5ex\hbox{$\; \buildrel < \over \sim \;$}}
\def\simgt{\lower.5ex\hbox{$\; \buildrel > \over \sim \;$}}

\def\beq{\begin{equation}}
\def\eeq{\end{equation}}
\def\tot{\mathrm{tot}}
\def\dm{\mathrm{dm}}
\def\br{\mathrm{b}}
\def\bc{_\mathrm{bc}}
\def\del{\delta}

\newcommand{\beqa}{\begin{eqnarray}}
\newcommand{\eeqa}{\end{eqnarray}}

\newcommand{\rme}{{\rm e}}
\newcommand{\vbc}{v_{\rm bc}}
\renewcommand{\vec}[1]{{\bf #1}}

\title{Suppression and Spatial Variation of Early Galaxies and Minihalos}

\author[Tseliakhovich, Barkana \& Hirata]
{Dmitriy Tseliakhovich$^{1}$\thanks{E-mail: dimlyus@caltech.edu},
Rennan Barkana$^2$ \& Christopher M. Hirata$^3$\\ $^{1}$California
Institute of Technology, M/C 249-17, Pasadena, California 91125, USA\\
$^{2}$ Raymond and Beverly Sackler School of Physics and Astronomy,
Tel Aviv University, Tel Aviv 69978, Israel \\ $^{3}$ California
Institute of Technology, M/C 350-17, Pasadena, California 91125, USA}

\date{December 12, 2010}

\begin{document}

\maketitle

\label{firstpage}

\begin{abstract}
We study the effect of the relative velocity of dark matter and
baryonic fluids after the epoch of recombination on the evolution of
the first bound objects in the early universe. Recent work has shown
that, although relative motion of the two fluids is formally a second
order effect in density, it has a dramatic impact on the formation and
distribution of the first cosmic structures. Focusing on the gas
content, we analyze the effect of relative velocity on the properties
of halos over a wide range of halo masses and redshifts. We calculate
accurately the linear evolution of the baryon and dark matter
fluctuations, and quantify the resulting effect on halos based on an
analytical formalism that has been carefully checked with simulations
in the case with no relative velocity. We estimate the effect on the
abundance of and gas fraction in early halos. We find that the
relative velocity effect causes several changes: (i) the
characteristic mass that divides gas-rich and gas-poor halos is
increased by roughly an order of magnitude, from $2\times
10^4\,M_\odot$ to about $2\times 10^5\,M_\odot$; (ii) this
characteristic mass has a large scatter (full width at half maximum is $\sim 1.5 \times 10^5 M_{\odot}$ at $z=20$); (iii) the
fraction of baryons in star-less gas minihalos is suppressed by a
factor of 4 at $z=20$; (iv) the fraction of baryons in halos that can
cool and form stars is suppressed by a factor of 1.5 at $z=20$; and
(v) there are enhanced spatial variations of these various fractions.
\end{abstract}


\section{Introduction}

One of the most important questions in astrophysics today is
understanding the formation and evolution of the first bound
structures. Significant theoretical and observational efforts are
devoted to understanding the properties of the first galaxies and
minihalos, at what redshifts they form and how they influence the
epoch of reionization.  Observations, most notably of the cosmic
microwave background (CMB), have established the basic parameters for
the initial conditions for structure formation \citep{WMAP1}, thus
providing a foundation for theoretical work on the first structures.
Advances in computation have made it possible to simulate the
formation of the first stars
\citep{abel, OShea05, Yoshida08}.  Meanwhile, several efforts are 
underway to probe the structure of the intergalactic medium (IGM)
during the reionization epoch using the 21 cm line of hydrogen, and
second-generation experiments may be able to explore the early stages
of reionization.

To answer these and many other questions it is imperative to know the
correct initial conditions that led to the formation of the first
bound objects and to account for all subtle effects that influence
evolution of the density perturbations after recombination.  The
critical role of initial conditions has been discussed by \citet{NYB},
who showed that three commonly used setups lead to significantly
different abundances and properties of the first star-forming gas
clouds as well as first gas-rich minihalos.


There are two major classes of early-type objects that must be
analyzed. The first class consists of large halos in which the gas can
cool and form stars; these are the presumed sites of the first dwarf
galaxies, which represent the first source of metals in the Universe,
and provide ultraviolet photons that begin the decoupling of the
hydrogen spin temperature from the CMB \citep{MMR} and eventually
start the epoch of reionization.  The second class consists of smaller
halos (``minihalos'') that are too small for molecular cooling, but
still affect the epoch of reionization by acting as a sink for
ionizing photons \citep{Haiman01, BL02, Iliev05, Ciardi05} and may
generate a 21 cm signal from collisional excitation of H{\sc\,i}
\citep[e.g.][]{Iliev03, FO06}.  
It is important to understand both the abundance and distribution of
halos, as well as the precise boundaries separating halos that undergo
cooling and star formation, those that collect baryons in their
potential wells but do not cool, and the lightest halos that exist
only as dark matter structures and do not collect gas.

An important effect that was previously overlooked is that of the
relative velocity of dark matter and baryonic fluids
\citep{TH10}. This effect leads to power suppression on scales that
correspond to the first bound halos between $10^4 \ M_{\odot}$ and
$10^8 \ M_{\odot}$ and delays the formation of the first objects. More
importantly this effect introduces scale-dependent bias and
stochasticity, leading to significant qualitative changes in the
distribution of the first objects. The relative velocity effect is
especially important on the small scales where the first stars and
galaxies form.  Introduction of this effect dramatically changes the
gas distribution inside the first halos and changes the characteristic
mass of gas-rich objects. \citet{Dalal10} recently calculated
analytically the effect on the gas content of halos and found a large
effect on the fluctuations of the Lyman-$\alpha$ background at high
redshifts. Their analysis, however, was based on a very simplified model of
which halos can form stars and in what abundance. In this paper we
carry out a detailed analytical study of the distribution of gas and
stars in the first halos. 

The rest of the paper is organized as follows.  Section~\ref{S:IC}
reviews the relative velocity effect (Sec.~\ref{sec1}) and improves
the analysis of \citet{TH10} to account for spatial variation of the
sound speed (Sec.~\ref{sec2}).  Section~\ref{S:FH} then investigates
the early halos and their gas content, focusing on computation of the
filtering mass (Sec.~\ref{sec3}) and then examining the fraction of
baryons in minihalos and in larger halos that can cool, including an
analysis of spatial variations in the baryon budget (Sec.~\ref{sec4} and ~\ref{sec5}).
We summarize our results in Sec.~\ref{sec6} and compare them to other
recent work.

The numerical results and plots shown in this paper assume a cosmology with present-day baryon density $\Omega_{\rm b,0}=0.044$, CDM density $\Omega_{\rm c,0}=0.226$, dark energy density $\Omega_{\rm 
\Lambda,0}=0.73$, Hubble constant $H_0=71$ km$\,$s$^{-1}\,$Mpc$^{-1}$, and adiabatic primordial perturbations of variance $\Delta^2_\zeta(k_\star)=2.42\times 10^{-9}$ at $k_\star = 0.002$ Mpc$^{-1}$ and with slope $n_s=0.96$.

\section{Initial conditions for halo formation}
\label{S:IC}

In this section we detail the formalism necessary for generation of
correct initial conditions taking into account two important effects
that are often overlooked in the literature. First of all we introduce
the effect of relative velocity of dark matter and baryonic fluids
after recombination. This effect, first studied in \citet{TH10}, is
nominally a second order effect in the perturbation theory of density
evolution and hence has been ignored in studies based on the linear
theory. Secondly, we emphasize the importance of a correct treatment
of the sound speed variations in the time between recombination and $z
\sim 200$ due to residual Compton heating of the electrons on the CMB
photons. As we show later in the paper, both effects play a
significant role during the epoch of first halo formation and
dramatically impact gas fractions in the first bound objects.

\subsection{Relative velocity of dark matter and baryonic fluids}
\label{sec1}

Before recombination, the baryons are tightly coupled to the photons
and the sound speed is highly relativistic, $\sim c/\sqrt{3}$. As the
universe expands and cools the electrons recombine with the protons
and the universe becomes transparent, leading to a kinematic
decoupling of the baryons from the radiation around $z_{\rm dec}
\approx 1000$. After recombination the sound speed of the baryons drops
precipitously down to thermal velocities, whereas the dark matter
velocity remains unaffected, and after recombination the dark matter
motion with respect to the baryons becomes supersonic. The relative
velocity can be written as:
\beq
{\bf v}_{\rm bc}({\bf k}) = \frac{\hat{\bf k}}{ik}[\theta_{\rm b}({\bf
k}) - \theta_{\rm c}({\bf k})],
\eeq
where $\hat{\bf k}$ is a unit vector in the direction of ${\bf k}$,
and $\theta \equiv a^{-1}\nabla\cdot{\bf v}$ is the velocity
divergence (we use comoving coordinates).

The variance of this relative velocity is
\beqa \label{eq:vbc}
\langle v_{\rm bc}^2(\vec{x}) \rangle &=& \int \frac{dk}{k} \Delta_\zeta^2(k)\left[ \frac{\theta_b (k) - \theta_c (k) }{k}\right]^2
\nonumber \\
&=& \int \frac{dk}{k} \Delta^2_{\rm vbc}(k),
\eeqa
where $\Delta_\zeta^2(k) = 2.42\times10^{-9}$ is the initial curvature
perturbation variance per $\ln k$. Integration of Eq.~(\ref{eq:vbc})
at the time of recombination\footnote{Technically the effective
redshift of kinematic decoupling \citep{EH}, since recombination is an
extended process.} ($z_{\rm rec} = 1020$) shows that the dark matter
moves relative to the baryons with root-mean square velocity
$\sim30\,$km$\,$s$^{-1}$ corresponding to a Mach number of ${\cal
M}\equiv v_{\rm bc}/c_{\rm s}\sim 5$.  This supersonic relative motion
allows baryons to advect out of the dark matter potential wells and
significantly suppresses the growth of structure at wave numbers
higher than
\beq
k_{\rm vbc} \equiv \left.\frac{aH}{\langle v_{\rm bc}^2\rangle^{1/2}}\right|_{\rm dec} 
= \frac{k_{\rm J}}{\cal M}
\sim 40\,{\rm Mpc}^{-1},
\label{eq:kvbcdef}
\eeq
were $k_{\rm J}$ is the Jeans wave number.

As shown in \citet{TH10} the relative velocity of the baryons and cold
dark matter (CDM) is coherent over scales of several comoving Mpc and
the velocity in each coherence region is well-described by a
3-dimensional Gaussian probability distribution with variance
\beq
\sigma^2_{\rm vbc} = \left\langle \left| {\bf v}_{\rm bc}({\bf x}) 
\right|^2\right\rangle.
\eeq
(Note that this is the {\em total} variance, i.e., including
velocities in all 3 directions; the variance per axis is smaller by a
factor of 3.)

To see how the relative motion of baryons and dark matter affect the
formation of the first objects we need to solve a system of evolution
equations that incorporate this effect. The system of equations
describing a high-$k$ perturbation mode in the presence of a
background relative velocity is
\beqa
\frac{\partial \delta_{\rm c}}{\partial t} &=&
\frac ia{\bf v}^{\rm(bg)}_{\rm bc}\cdot{\bf k} \delta_{\rm c}
 - \theta_{\rm c},
\nonumber \\
\frac{\partial \theta_{\rm c}}{\partial t} &=&
\frac ia{\bf v}_{\rm bc}^{\rm(bg)}\cdot{\bf k}\theta_{\rm c}
 -\frac{3H^2}{2}
(\Omega_{\rm c}\delta_{\rm c} + \Omega_{\rm b}\delta_{\rm b}) 
- 2H\theta_{\rm c},
\nonumber \\
\frac{\partial \delta_{\rm b}}{\partial t} &=& -\theta_{\rm b}, {\rm ~and}
\nonumber \\
\frac{\partial \theta_{\rm b}}{\partial t} &=& -\frac{3H^2}{2}
(\Omega_{\rm c}\delta_{\rm c} + \Omega_{\rm b}\delta_{\rm b}) 
- 2H\theta_{\rm b}
 + \frac{c_{\rm s}^2k^2}{a^2} \delta_{\rm b}.
\label{eq:evoleqn}
\eeqa

The $\vbc$ terms are nominally second order in perturbation theory,
and hence one may wonder why they, rather than other second-order
terms, are included.  The reason is that the expansion parameter for
these terms is not the density perturbation $\delta$, but rather the
ratio of the advection terms (e.g. $v_{\rm bc}^{\rm(bg)}k \delta_c/a$
in the $\delta_c$ equation) to the linear terms
(e.g. $\partial\delta_c/\partial t\sim\delta_c/H$).  This ratio is
\beq
\frac{v_{\rm bc}^{\rm(bg)}k}{aH}.
\eeq
One can see that this expansion parameter increases as one goes to
smaller scales and is of order unity at $k \sim k_{\rm vbc}$.  Thus
the $\vbc$ terms become nonperturbative at small scales $k>k_{\rm
vbc}$, and when treating these small scales one must keep these terms
even if they are formally higher order in the perturbation theory.

\subsection{Complete heating model}
\label{sec2}

The system of equations of Eq.~(\ref{eq:evoleqn}) assumes a spatially
uniform sound speed which is a good first-order
approximation. However, as shown in \citet{NB05}, it underestimates
baryon density fluctuations by more than 30 percent at $z = 100$ and
10 percent at $z = 20$ for large wavenumbers.  A fully correct
treatment of baryon density evolution requires analysis of the Compton
heating from the CMB on the sound speed and fluctuations in the
temperature distribution. Following \citet{NB05}, we re-write the
sound speed term of the last equation of Eq.~(\ref{eq:evoleqn}) as
\beq
\frac{c_{\rm s}^2k^2}{a^2} \delta_{\rm b} \rightarrow \frac{k^2}{a^2}\frac{k_B \bar{T}}{c^2 \mu m_H} (\delta_{\rm b} + \delta_{\rm T}),
\label{Eq:cs}
\eeq
where $\delta_{\rm T}$ is the temperature perturbation which evolves as:
\begin{equation}
\label{gamma} \frac{d \delta_T} {d t} = \frac{2}{3} \frac{d
\delta_\br} {dt} + \frac{x_e(t)} {t_\gamma}a^{-4} \left\{
\delta_\gamma\left( \frac{\bar{T}_\gamma}{\bar{T}} -1\right)
+\frac{\bar{T}_\gamma} {\bar{T}} \left(\delta_{T_\gamma} -\delta_T
\right) \right\}.
\end{equation}
The second term on the right-hand side accounts for the Compton
scattering of the CMB photons on the residual electrons from
recombination. Here $x_e(t)$ is the electron fraction relative to the
total number density of gas particles\footnote{This is different from
the recombination literature, which often takes $x_e$ to be normalized
to the number of hydrogen nuclei.  At low redshifts these differ by 8
per cent due to the presence of helium.}, $\bar{T}_\gamma=[2.725\ {\rm
K}]/a$ is the mean CMB temperature, and
\begin{equation}
\label{tgamma} t_\gamma^{-1} \equiv \frac{8} {3} \rho_\gamma^0
\frac{\sigma_{T}\, c} {m_e} = 8.55 \times 10^{-13} {\mathrm{yr}}^{-1},
\end{equation}
where $\sigma_{\rm T}$ is the Thomson scattering cross section,
$\rho_\gamma^0$ is the photon energy density at $z=0$, and $\bar{T}$
is the average temperature of the baryons, which can be calculated
using the first law of thermodynamics:
\begin{equation}
\label{mean} \frac{d \bar{T}} {dt} = - 2 H \bar{T} +
\frac{x_e(t)}{t_\gamma}\, (\bar{T}_\gamma - \bar{T})\, a^{-4}.
\end{equation}
Accounting for Compton heating of the residual electrons by the CMB
photons is especially important on small scales ($k > 1$ Mpc$^{-1}$),
which are also impacted by the relative motion effect.

\section{First halos and their gas content}
\label{S:FH}

Both of the effects discussed above have a significant impact on the
evolution of density perturbations on small scales and affect the
formation of the first dark matter halos, as well as the subsequent
accretion of the baryons and the formation of the first stars. We
investigate the specific effects by studying the change in the
characteristic mass scale that divides gas-rich and gas-poor halos
produced by the relative velocity of the dark matter and baryonic
fluids.

\subsection{Filtering mass}
\label{sec3}

In the $\Lambda$CDM universe, virialized dark matter halos form
hierarchically on extremely small scales at very early times and start
accreting baryons into their potential wells. If halos are heavy
enough, accretion proceeds to the point where baryons start cooling
through molecular line emission, condensing into the first stars and
galaxies. This accretion is counteracted by the bulk motion of baryons
with respect to dark matter as well as by the thermal gas
pressure. The combination of the two effects leads to the presence of
the minimal halo mass scale at which baryons are still able to
effectively accrete onto a halo.

To study the effect of halo formation and baryonic accretion it is
convenient to divide space into a large number of patches of the size
of the relative velocity coherence scale. In each patch with a given
mean density and bulk velocity, we follow the evolution of density
perturbations including the spatial variation of the baryonic speed of
sound due to Compton heating from the CMB \citep{NB05}. 

By evolving the system of equations~(\ref{eq:evoleqn}) with the
correct sound speed term of Eq.~(\ref{Eq:cs}) in each patch, we
calculate the baryonic and dark matter power spectra. Their ratio is
constant on large scales (small $k$), and drops at high $k$ due to the
suppression of growth by the baryonic pressure.
\citet{gh98} originally defined a ``filtering'' scale (essentially 
a time-averaged Jeans scale) that they used to identify the largest
scale on which the baryon fluctuations are substantially suppressed
compared to those of the dark matter. We use the generalized
definition from
\citet{NB07}, in which the baryon-to-total ratio is expanded to linear
order in $k^2$, and written in the following form:
\beq
\frac{\delta_\br}{\delta_\tot}=1-\frac{k^2}{k_F^2}+r_{\rm LSS},
\label{Eq:Ratio3}
\eeq
where the total density perturbation
$\delta_\tot=f_\br\delta_\br+f_\dm\delta_\dm$ (in terms of the cosmic
baryon and dark matter mass fractions $f_\br$ and $f_\dm$), and the
$k$-independent $r_{\rm LSS}$ term (which is negative) describes the
relative baryon-to-total difference in the limit of large scale
structure, i.e., where both the $v_{\rm bc}$ effect and the thermal
pressure of the gas are negligible (and restricted also to scales
below the baryon acoustic oscillations).

\begin{figure}
\includegraphics[width=3.4in]{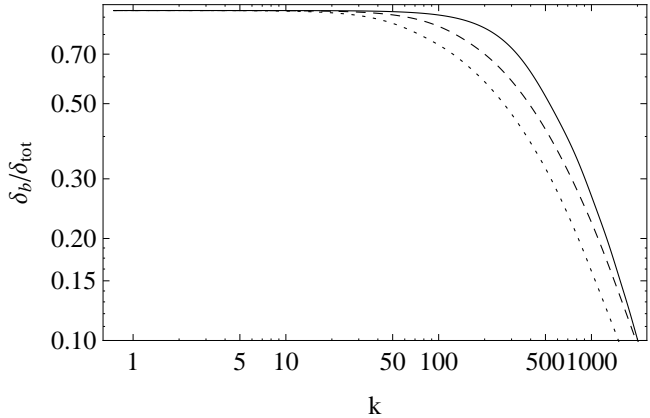}
\caption{\label{fig:Ratio3aver} Perturbation 
ratio $\delta_b/\delta_{\rm tot}$ vs. comoving wavenumber $k$
evaluated at $z = 20$ for the cases of $\vbc = 0$ (solid curve), $\vbc
= 1\sigma_{vbc}$ (dashed curve), and $\vbc = 2\sigma_{vbc}$ (dotted
curve). In all cases overdensities are isotropically averaged over the
direction of $\vec{k}$ with respect to $\vec{\vbc}$.}
\end{figure}

In Fig.~\ref{fig:Ratio3aver} we plot the isotropically averaged
perturbation ratio $\delta_b/\delta_{\rm tot}$ by averaging over the
direction of $\vec{k}$ with respect to $\vec{\vbc}$. On large scales
the ratio is very close to constant, and using Eq.~(\ref{Eq:Ratio3})
we can deduce $r_{\rm LSS} = -0.054$ at $z=20$. The filtering scale $k_F$
is obtained by fitting Eq.~(\ref{Eq:Ratio3}) to the calculated values
of the ratio $\delta_b/\delta_{\rm tot}$ from
Fig.~\ref{fig:Ratio3aver}. This allows us to define the filtering mass
in terms of the filtering wavenumber:
\begin{equation}
M_F=\frac{4\pi}{3}\bar{\rho_0}\left(\frac{\pi}{k_F}\right)^3, \label{Mf}
\end{equation}
where $\bar{\rho_0}$ is the mean matter density today. We note that
this relation is $\frac18$ of the definition originally used by
\citet{gnedin00}, who also used a non-standard definition of the Jeans
mass.

The filtering mass plays an extremely important role in understanding
the evolution of the first halos, as it provides a good approximation
for the boundary between the gas-rich halos and halos that do not
contain substantial quantities of gas. Traditionally one would assume
that the separation between gas-rich and gas-poor objects is
represented by the Jeans scale, which is the minimum scale on which a
small gas perturbations will grow due to gravity overcoming the
pressure gradient.  However, the Jeans scale is related only to the
evolution of the perturbations at a given point in time and does not
account for significant variation of the Jeans mass with time.  The
filtering mass on the other hand reflects the baryonic pressure
effects integrated over the entire history of the Universe, and
provides a much better approximation to the boundary between gas-rich
and gas-poor halos.

\begin{figure}
\includegraphics[width=3.4in]{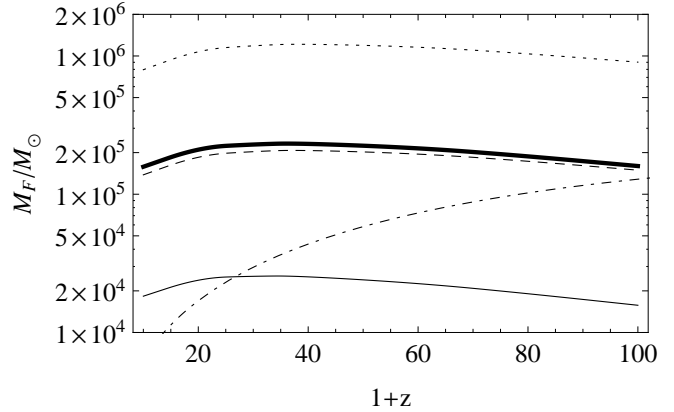}
\caption{\label{Fig:MFz} Evolution of the filtering mass with 
redshift in the regions with $\vbc = 0$ (thin solid line), $\vbc =
1\sigma_{vbc}$ (dashed line), $\vbc = 2\sigma_{vbc}$ (dotted line) and
global average case (thick solid line). We also show the evolution of
the Jeans mass $M_J$ (dot-dashed line).}
\end{figure}

An extensive study of the filtering mass properties and evolution
history in models without the relative velocity effect was performed
in \citet{NB05} and \citet{NB07}. The properties of the filtering
mass, however, are significantly modified in the regions where the
bulk motion of baryons with respect to dark matter potential wells is
significant. In regions with high values of $\vbc$ baryons tend to
advect out of the collapsing dark matter halos, significantly
increasing the filtering mass. We demonstrate this in
Fig.~\ref{Fig:MFz} where we plot the evolution of the filtering mass
with redshift in the regions with $\vbc/\sigma_{\rm vbc} = 0$, 1, and
2. We also show the globally averaged case by integrating the
filtering mass over the full probability distribution of the relative
velocity, given by:\footnote{This is the distribution of
the magnitude of $\vbc$, where the vector ${\bf v}_{\rm bc}$ is the
result of linear perturbations and hence is drawn from a multivariate
Gaussian.  It thus happens to be the same as the Maxwell-Boltzmann
distribution, even though the bulk velocities of baryons have nothing
to do with thermal motions of particles.}
\beq
P_{\rm vbc}(v) = \left(\frac{3}{2\pi\sigma_{\rm vbc}^2}\right)^{3/2} 4\pi v^2 \exp\left(-\frac{3v^2}{2\sigma_{\rm vbc}^2}\right).
\label{vbcPDF}
\eeq
As noted earlier, the variance per axis is $\sigma_{\rm vbc}^2/3$.

In Fig.~\ref{Fig:MFz} we also compare the filtering mass with the
Jeans mass defined as:
\beq
M_J = \frac{4 \pi}{3} \bar{\rho_0} \left(\frac{\pi}{k_J}\right)^3,
\eeq
where $k_J = \sqrt{2/3}a H/c_{\rm s}$ is the Jeans scale (defined by
setting the right-hand side of Eq.~(\ref{eq:evoleqn}) to zero, without
the relative velocity term, and neglecting here the correction of
Eq.~(\ref{Eq:cs})). Fig.~\ref{Fig:MFz} shows that the filtering mass
reaches a maximum value at redshift $z \sim 40$ (and generally varies
only slightly throughout the plotted redshift range), whereas the
Jeans mass continuously decreases with time due to the drop in the
sound speed of the gas as the Universe cools.

\begin{figure}
\includegraphics[width=3.4in]{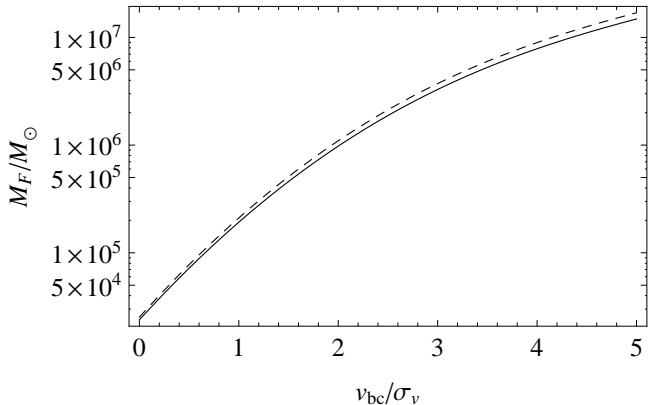}	
\caption{\label{Fig:MofV1} This figure shows the filtering mass 
$M_F$ as a function of the relative velocity of the dark matter and
baryonic fluids at $z=20$ (solid line) and $z = 40$ (dashed line).}
\end{figure}

\begin{figure}
\includegraphics[width=3.4in]{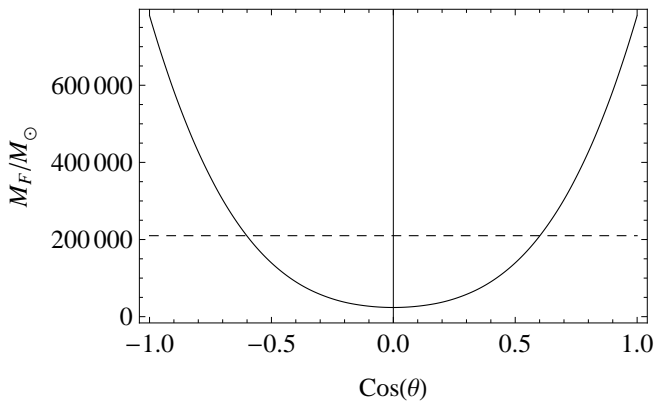}
\caption{\label{Fig:MofV2} The
lower panel shows the dependence of the filtering mass on the angle
$\theta$ between the directions of $\bf{v}_{bc}$ and the wavevector
$\mathbf{k}$ (solid line), and the isotropically averaged value of
$M_F$ (dashed horizontal line), in regions where $\vbc =
2\sigma_{vbc}$ at $z = 20$.}
\end{figure}

The filtering mass represents a time averaged Jeans mass and hence it
decreases at the low redshifts, however, right after recombination
baryonic perturbations on small scales are highly suppressed and they
only catch up gradually, causing the filtering mass to increase from
low initial values. We emphasize that in the regions with a high value
of the relative velocity the filtering mass is significantly larger
than in the regions with small values of $\vbc$ and hence the
formation of gas-rich objects in those regions proceeds quite
differently than in the regions with $\vbc \sim 0$. The filtering
scale and mass (from Eqs.~(\ref{Eq:Ratio3}) and (\ref{Mf})) in regions
with varying values of $\vbc$ are given in Table~\ref{Tab:Mfaver}, and
the filtering masses at $z=20$ and $z=40$ are plotted in Fig.~\ref{Fig:MofV1}. In Fig.~\ref{Fig:MofV2}, we
also show the dependence of the filtering mass on the angle $\theta$
between the direction of $\bf{v}_{bc}$ and that of the wavevector
$\mathbf{k}$ in regions where $\vbc = 2\sigma_{vbc}$ at $z = 20$; the
plot shows that the main contribution to the filtering mass comes from
the regions where the wavenumber $\mathbf{k}$ and the relative
velocity vector $\bf{v}_{bc}$ are parallel (or anti-parallel).

\begin{table}
\begin{center}
\begin{tabular}{l | c | c | r}
\hline
$\vbc/\sigma_{\vbc}$ & $P(>\vbc)$ & $k_F$ (Mpc$^{-1}$) & $M_F$ $(M_{\odot})$ \\ \hline
$4$ & $2.1\times 10^{-10}$ & $85$ & $7.75\times 10^6$ \\
$3$ & $5.9\times 10^{-6}$ & $113$ & $3.37\times 10^6$ \\
$2$ & $7.4\times 10^{-3}$ & $166$ & $1.07\times 10^6$ \\
$1$ & $0.392$ & $298$ & $1.85\times 10^5$ \\
$0$ & $1$ & $591$ & $2.39\times 10^4$ \\
\hline
\end{tabular}
\caption{\label{Tab:Mfaver} Filtering scale and filtering mass for the isotropic averaging of the direction of $\vec{k}$ with respect to $\vec{\vbc}$ 
at $z=20$.}
\end{center}
\end{table}

\subsection{Gas content of the first galaxies and minihalos}
\label{sec4}

We now investigate the amount of gas that falls into early haloes, and how much of this gas is capable of cooling.  Here we use analytical approximations -- namely the relation between the gas mass fraction $f_{\rm g}$ and filtering mass $M_F$, and the \citet{shetht99} mass function -- that have been calibrated against simulations with statistically isotropic initial conditions and no bulk relative velocity.  In our case with $v_{\rm bc}\neq0$ the power spectra are both reduced and slightly anisotropic, but we expect these approximations to still be a useful guide since statistical anisotropy (e.g. $\theta$-dependent filtering mass) can only appear at second order in scalars such as the halo mass function or gas content.

\begin{figure}
\includegraphics[width=3.4in]{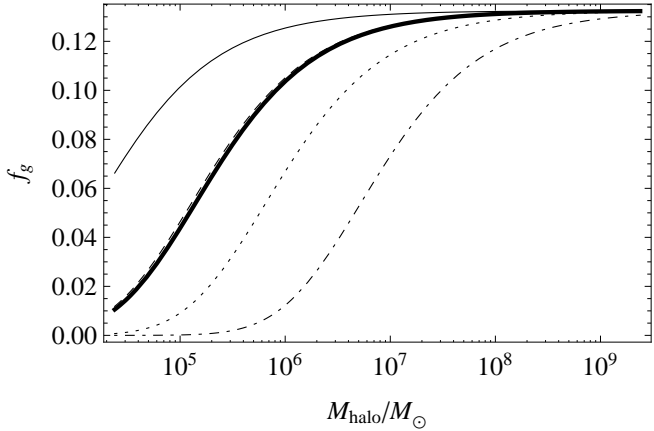}
\caption{\label{Fig:fgofM} Change in the gas fraction $f_g$ at 
$z = 20$ as a function of halo mass for regions with $\vbc = 0$ (thin
solid line), $\vbc = 1\sigma_{vbc}$ (dashed line), $\vbc =
2\sigma_{vbc}$ (dotted line), $\vbc = 4\sigma_{vbc}$ (dash-dotted
line) and isotropically averaged case (thick solid line, nearly
coincident with the dashed line).}
\end{figure}

There is no {\slshape a priori} reason to suppose that the filtering
mass, which is defined based on linear perturbations, can also
accurately describe properties of highly nonlinear, virialized
objects. Qualitatively we may argue that if pressure significantly
opposes gravity during the halo formation process (which for some time
is accurately described by linear theory) then it will significantly
suppress the amount of gas in the final virialized halo.
\citet{gnedin00} suggested based on simulations during cosmic 
reionization that the filtering mass accurately fits the mass for
which a halo contains half the mean cosmic baryon fraction $f_b$, and
fitted the simulation results to the following formula:
\begin{equation}
\label{f_g-alpha}
f_{\rm g}= f_{\br,0} \bigg[1+\left(2^{\alpha/3}-1
\right)\left(\frac{M_F}{M}\right)^\alpha \bigg]^{-3/\alpha},
\end{equation}
where $f_{\br,0}$ is the gas fraction in the high-mass limit. In this
function, a higher $\alpha$ causes a sharper transition between the
high-mass (constant $f_{\rm g}$) limit and the low-mass limit (assumed
to be $f_{\rm g} \propto M^3$).  This formula has subsequently been
found to agree with hydrodynamic simulations \citep{NBM,NYB} if we set
$\alpha \approx 0.7$ and $f_{\br,0}=f_\br (1+3.2 r_{\rm LSS})$
\citep{bl10}, and use the filtering mass as defined in Eq.~(\ref{Mf})
(which differs from \citet{gnedin00}, as noted earlier). Thus, at each
redshift, in each patch of the Universe we may calculate the local
value of $M_F$ and from it the gas fraction in halos of various total
mass.  In Fig.~\ref{Fig:fgofM}, we plot the gas fraction as a function
of halo mass in regions with varying values of relative velocity at $z
= 20$. It is clear that halos that would be gas-rich in the Universe
with no $\vbc$ effect become gas-poor in the regions where the
relative velocity is high. We also see that on average, introduction
of the $\vbc$ effect significantly lowers the gas fraction in all
halos with $M_h < 10^7 M_{\odot}$.

In order to find the total amount of gas in galaxies, we must
integrate over the halo mass function in each patch. We start with the
$\vbc=0$ case.  Standard models for halo formation are based on
spherical collapse calculations, in which the linear overdensity must
reach a critical threshold $\delta_c(z)$ for the corresponding region
to form a collapsed halo at redshift $z$. The halo abundance depends
on the statistics of fluctuations on various scales, which can be
parameterized by the function $S(R)$, the variance of fluctuations in
spheres of radius $R$ ($S$ is also a function of redshift). In general
we define $f(\del_c(z),S)\, dS$ to be the mass fraction contained at
$z$ within halos with mass in the range corresponding to $S$ to
$S+dS$. We convert between halo mass $M$ and (initial comoving) radius
$R$ using the cosmic mean density. The halo abundance is then
\beq 
\frac{dn}{dM} = \frac{\bar{\rho}_0}{M} \left|\frac{d S}{d M} 
\right| f(\del_c(z),S), \label{eq:abundance} 
\eeq 
where $dn$ is the comoving number density of halos with masses in the range $M$ to $M+dM$. In the model of \citet{ps74}, 
\beq 
f_{\rm PS}(\del_c(z),S) = \frac{1} {\sqrt{2 \pi}} \frac{\nu }{S} \rme^{-\nu^2/2}, \label{eq:PS} 
\eeq 
where $\nu=\del_c(z)/\sqrt{S}$ is the number of standard deviations that the critical collapse overdensity represents on the mass scale $M$ 
corresponding to the variance $S$.

However, the Press-Schechter mass function fits numerical simulations only qualitatively, and in particular it substantially underestimates the 
abundance of the rare halos that host galaxies at high redshift. The halo mass function of \citet[]{shetht99}, which fits numerical simulations much 
more accurately, is given by
\beq 
f_{\rm ST}(\del_c(z),S) = A' \frac{\nu }{S} \sqrt{\frac{a'} {2
\pi}} \left[ 1+\frac{1}{(a' \nu^2)^{q'}} \right] \rme^{-a'\nu^2/2},
\label{eq:ST} 
\eeq 
with best-fit parameters \citep{shetht02} $a'=0.75$ and $q'=0.3$, and where the normalization to
unity is ensured by taking $A'=0.322$.

The critical density of collapse $\del_c(z)$ is independent of mass
and equals 1.69 in the Einstein de-Sitter limit, valid over a wide
range of redshifts. Its value decreases at low redshift due to the
cosmological constant, but more relevant for this paper is the
decrease at very high redshift due to the effects of the baryons and
radiation. The decrease is by $\sim 0.05(1+z)$ per cent from the
Einstein de-Sitter value \citep{NB07}, a small effect that is however
greatly amplified by the fact that galaxies at the highest redshifts
form in halos that correspond to very rare density fluctuations.

\begin{figure}
\includegraphics[width=3.4in]{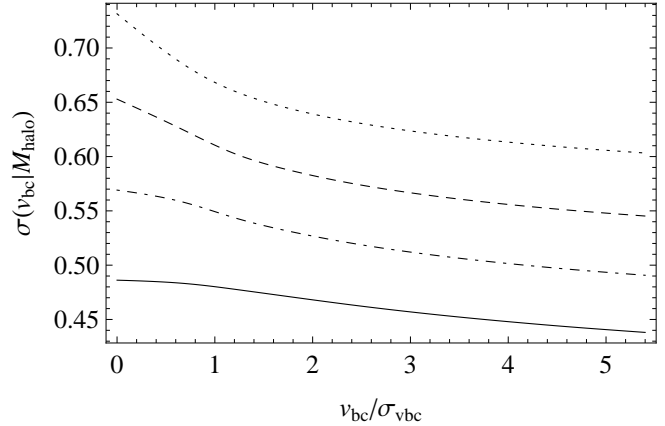}
\caption{\label{Fig:SigofV} Dependence of $\sigma(\vbc|M_h)$ on 
the relative velocity $\vbc$ at $z=20$ for a fixed mass of the
collapsed halos: $M_h = 10^7 M_\odot$ (solid line), $M_h = 10^6
M_\odot$ (dot-dashed line), $M_h = 10^5 M_\odot$ (dashed line), and
$M_h = 10^4M_\odot$ (dotted line).}
\end{figure}

In our approach we must calculate how the halo mass function varies in
different regions. The Press-Schechter model has been extended
\citep{bond91} to describe the variation of the halo abundance in
regions of various density, and we can generalize this to include the
bulk velocity by including the variation of the function $S(R)$ with
$v\bc$. To demonstrate this dependence we plot in
Fig.~\ref{Fig:SigofV} $\sqrt{S}
\equiv \sigma(\vbc|M_h)$ as a function of the relative velocity at a 
fixed halo mass $M_h$. We see that the variance of the density
perturbations decreases with increasing $\vbc$, leading to a delay in
the collapse of dark matter halos. We also find that the change in
$\sigma$ is larger for halos of low mass since the power spectrum on
scales much larger than the filtering mass is unaffected by the
relative velocity.

\begin{figure}
\includegraphics[width=3.4in]{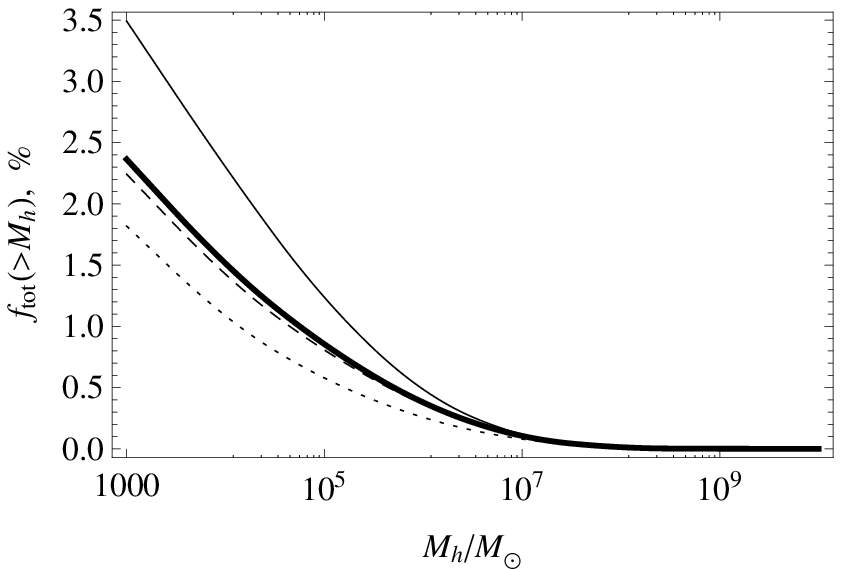}
\includegraphics[width=3.4in]{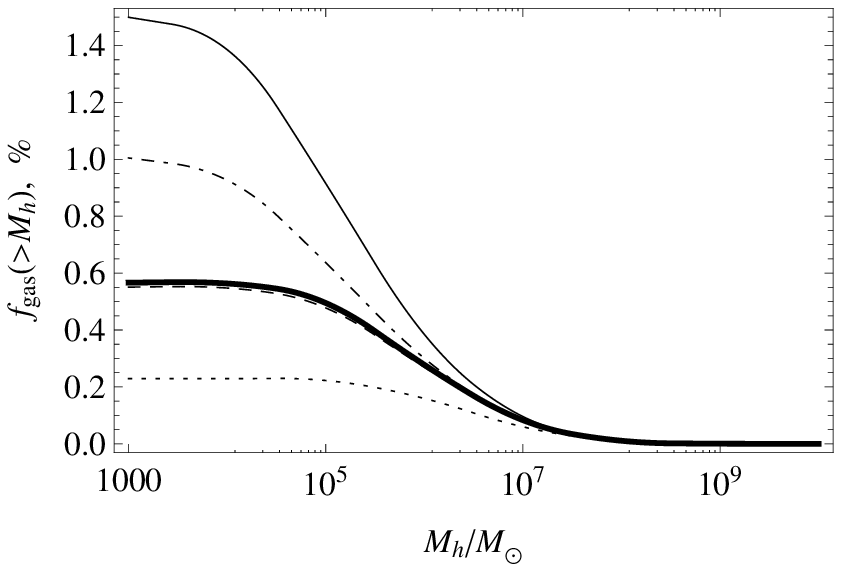}
\caption{\label{Fig:frac1} The mass fraction in halos 
above $M_h$ (upper panel) and the gas fraction in halos (lower panel)
at redshift $z=20$ for the case $\vbc/\sigma_{\rm vbc} = 0$ (thin
solid line), 1 (dashed line), 2 (dotted line), and for the globally
averaged case (thick solid line). In the lower panel we also show the
case where we fix the value of $M_F$ as calculated for $\vbc = 0$ and use the correct globally averaged halo mass function
including the variation with $\vbc$ (dot-dashed line).}
\end{figure}

We can summarize the important effects of the change in halo abundance
and halo gas content at a given mass scale by calculating various gas
fractions at each redshift. This can be done by using the Sheth-Tormen
mass function and accounting for the changes due to the relative
velocity effect. We calculate the fraction of the total matter density
in halos above a certain mass scale by
\beq
f_{\rm tot}( > M_h) = \int_{M_h}^{\infty}\frac{M}{\bar{\rho}_0}
\frac{dn}{dM}dM\ ,
\eeq
and the fraction of the baryon density contained in those halos using
Eq.~(\ref{f_g-alpha}):
\beq
f_{\rm gas}( > M_h) = \int_{M_h}^{\infty}\frac{M}{\bar{\rho}_0}
\frac{dn}{dM}\frac{f_g}{f_b}dM\ .
\eeq
We plot both fractions in Fig.~\ref{Fig:frac1} at $z=20$ for
$\vbc/\sigma_{\rm vbc} = 0$, 1, and 2, and for the globally averaged
case, where we take into consideration the global distribution of the
$\vbc$. The plot clearly shows that in regions with high relative
velocity the gas fraction in halos is dramatically suppressed. The
global average (which comes out very close to the $\vbc/\sigma_{\rm
vbc} = 1$ case) gives a suppression by a factor of 2.7 of the total
gas fraction in halos. In order to separate out the various effects,
we plot one case in which we use the correct halo mass function (as it
varies with $\vbc$) but fix the filtering mass to the $\vbc=0$
value. We find that the suppression arises from comparable
contributions from the change in halo numbers (about a factor of 1.8)
and from the reduction in the internal halo gas fractions (about a
factor of 1.5). 

Stars are understood to form at high redshift out of gas that cooled
and subsequently condensed to high densities in the cores of dark
matter halos. Since metals are absent in the pre-stellar universe, the
earliest available coolant is molecular hydrogen (H$_2$), and thus the
minimum halo mass that can form a star is set by requiring the
infalling gas to reach a temperature of several hundred Kelvin
required for exciting H$_2$ to the $J\ge 2$ rotational levels
\citep{th2}. This has been confirmed with high-resolution numerical
simulations containing gravity, hydrodynamics, and chemical processes
in the primordial gas \citep{abel,bromm,fuller,Yoshida,reed}. These
simulations imply a minimum halo circular velocity $V_c \sim 4.5$ km
s$^{-1}$ for forming a star, where $V_c=\sqrt{GM/R}$ in terms of the
halo virial radius $R$. The simulations show that in a halo above the
minimum mass (which at $z=20$ is $M_{\rm min} \approx 6\times 10^5
M_{\odot}$), the gas cools in the dense center and forms at least one
star very quickly; this is understood theoretically since both the
cooling time and the dynamical time are a small fraction of the cosmic
age at that time. We are thus interested in the total gas fraction in
halos above this cooling threshold; if there is a fixed star formation
efficiency in these halos, then the this gas fraction is directly
proportional to the stellar density in each region.

\begin{figure*}
\centering
\includegraphics[width=3.4in]{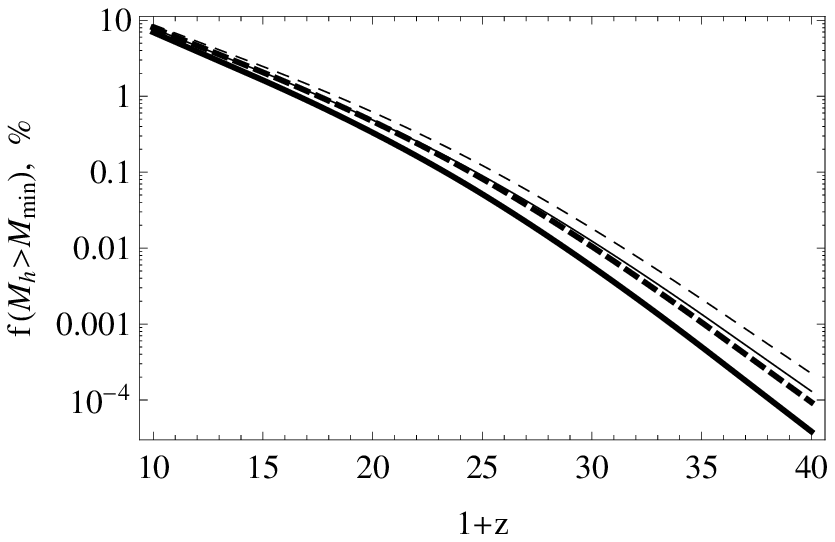}
\includegraphics[width=3.4in]{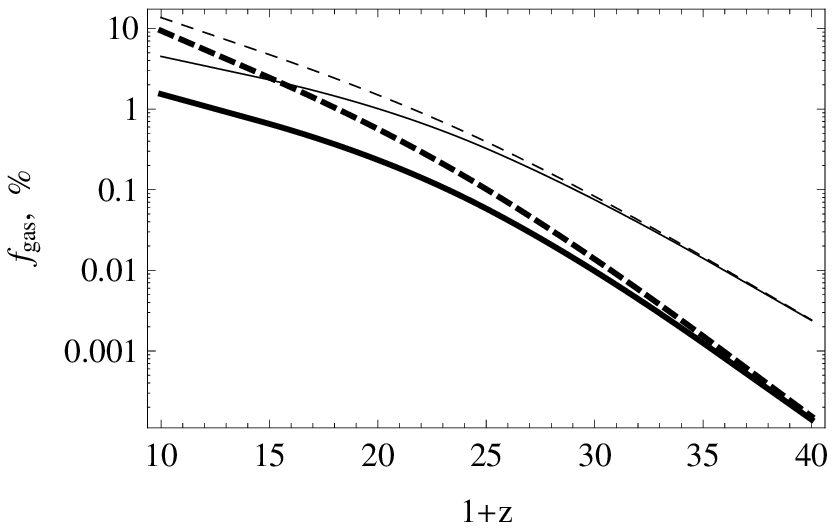}
\caption{\label{Fig:fracofz} In the left panel we plot the total mass fraction in halos 
above the cooling mass (dashed lines) and total gas fraction in halos
above the minimum cooling mass (solid lines). In the right panel we show total gas fraction in halos (dashed lines) and gas fraction in minihalos, i.e., in
halos below minimum cooling mass (solid lines).  All plots show
two cases: no $\vbc$ (regular lines) and the correct case where the
$\vbc$ effect is taken into account and the isotropic averaging is
performed (thick lines).}
\end{figure*}

We plot in Fig.~\ref{Fig:fracofz} the evolution of various gas and
total mass fractions as a function of redshift. Even without the
relative velocity effect, there is some (spatially-uniform)
suppression predicted for the gas fraction in halos that can cool
(i.e., a suppression of the overall star formation) by a factor of 1.2
at $z=20$ and 1.5 at $z=40$ (relative to the cosmic baryon fraction);
this is due to the fact that the baryon perturbations are still
catching up to the dark matter perturbations at these redshifts, even
on large scales (beyond the filtering mass), and simulations suggest
that non-linear halo formation amplifies the remaining differences
\citep{NYB,bl10}. The relative velocity effect adds an additional
suppression of cosmic star formation by a factor of 1.6 at $z=20$ and
3.4 at $z=40$. The relative velocities have a larger effect on the gas in
minihalos, the smaller halos that accrete gas that cannot cool. Since
the total mass fraction in halos continues to increase as we consider
smaller and smaller halos masses (Fig.~\ref{Fig:frac1}), the total
amount of gas in minihalos is very sensitive to the filtering mass,
which is what produces the (gradual) low-mass cutoff in gas accretion
onto halos. In the absence of the relative velocities, the total gas
fraction in halos at $z=20$ is $1.5\times 10^{-2}$, consisting of $1.0\times 10^{-2}$ in
minihalos and $5\times 10^{-3}$ in galaxies.  At $z=40$, these gas fractions are
$2.4\times 10^{-5}$, $2.3\times 10^{-5}$, and $1\times 10^{-6}$, respectively. The relative
velocities, in the global average, reduce these fractions to $6\times 10^{-3}$,
$3\times 10^{-3}$, and $3\times 10^{-3}$ at $z=20$, and $1.5\times 10^{-6}$, $1.1\times 10^{-6}$,
and $4\times 10^{-7}$ at $z=40$. Note that the gas fraction above the
H$_2$ cooling mass is really an upper limit to the gas fraction that
undergoes star formation. Any significant feedback effect will raise
the effective threshold for star formation, making the total gas
fraction in halos correspond almost completely to star-less halos (see
discussion in Sec.~\ref{sec6}).

\subsection{Probability distribution functions}
\label{sec5}

In addition to plotting mean values of various quantities, it is
interesting to consider their variation in various regions. Above, we
have explicitly varied $\vbc$ but averaged over the density
fluctuations (note that the density and velocity are uncorrelated at a
given point). However, in order to calculate the full amount of
variation of various quantities, i.e., the probability distribution
function (PDF), we must explicitly vary both the mean density and
the value of $\vbc$ in each region.

In the absence of relative velocities, the extended Press-Schechter
model gives the variation of the Press-Schechter halo mass function in
regions of various mean density. No analytical generalization of this
formalism is known for the more accurate Sheth-Tormen model, but
\citet{BLflucts} suggested a hybrid prescription that adjusts the
abundance in various regions based on the extended Press-Schechter
formula, and showed that it fits a broad range of simulation results.
Generalizing this prescription to include the effect of relative
velocity, we set
\begin{eqnarray}
f_{\rm bias}(\del_c(z),\bar{\delta}_R,R,M,v\bc) \!\! \!\! &=& \!\!\!\!
\left[ \frac{f_{\rm ST}(\del_c(z),S')} {f_{\rm PS}
(\del_c(z),S')} \right] \nonumber
 \\ && \!\!\!\!
\times f_{\rm PS}
\left(\del_c(z)-\bar{\delta}_R,S'-S'(R)\right), \nonumber
\\ &&
\end{eqnarray}
where the mean overdensity in the patch is $\bar{\delta}_R$, and for a
given halo mass, the variance $S'$ is calculated using the power
spectrum modified by the local bulk velocity. The subtraction of
$S'(R)$ accounts for the fact that $\bar{\delta}_R$ arises from
density modes on scales larger than the patch size, leaving only the
remaining variance $S'-S'(R)$ for fluctuation modes within the patch
to supply the additional density needed to reach $\del_c(z)$ and thus
form a halo. In our case, the patches in which we will compute the
baryon collapse fraction PDF will be spheres of radius $R=3$ Mpc
(comoving).  Note that if used to compute a mass function, the above
formula gives the Lagrangian halo number density, while the Eulerian
density is larger by a factor of $1+\bar{\delta}_R$; however no such
transformation is necessary to compute the local {\em fraction} of gas
in halos.

\begin{figure}
\includegraphics[width=3.4in]{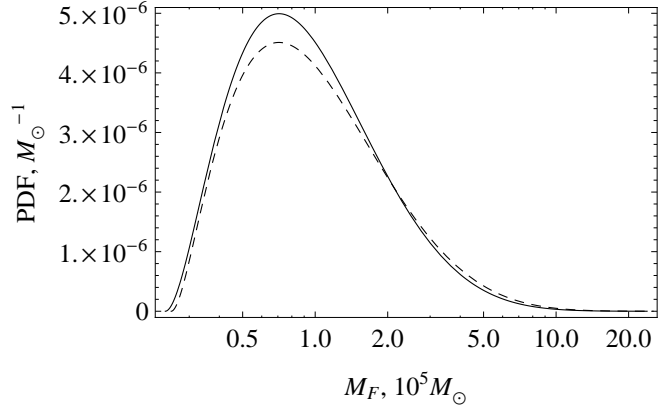}
\caption{\label{Fig:MFPDF}The PDF of the filtering mass $M_F$ at 
$z=20$ (solid line) and $z=40$ (dashed line). We consider cells of
radius $R = 3$ Mpc, and include the variation of $\vbc$ as well as the
mean density in each cell.}
\end{figure}

We start by calculating the PDF for the filtering mass $M_F$. In the
scenario without $\vbc$ we would have a universal value of $M_F$,
however, since various regions of space have different values of
relative velocity of baryonic and dark matter fluids this produces a
variation in $M_F$. The distribution of relative velocities is given
by Eq.~(\ref{vbcPDF}) and it translates into the distribution of $M_F$
using:
\beq
P_{M_F}(M_F) = P_{vbc}(\vbc) \frac{d\vbc}{dM_F}.
\eeq
The PDF of the filtering mass at $z=20$ and $z=40$ is plotted in
Fig.~\ref{Fig:MFPDF}. These distribution functions are essentially
determined by the distribution of the relative velocity and exhibit
clear peaks which correspond to values of the filtering mass around
the maximum of the $\vbc$ distribution, which occurs at $\vbc
\approx 1.2\sigma_{vbc}$. As noted before (Figs.~\ref{Fig:MFz} and 
\ref{Fig:MofV1}), the filtering mass does not vary much in this 
redshift range, but at $z=20$ it is slightly more sharply peaked while
the PDF at $z=40$ extends more towards high values of $M_F$. Filtering mass has a rather significant scatter with the full width at half maximum $\sim 1.5\times10^5 M_{\odot}$ at the redshift of $z=20$.

\begin{figure}
\includegraphics[width=3.4in]{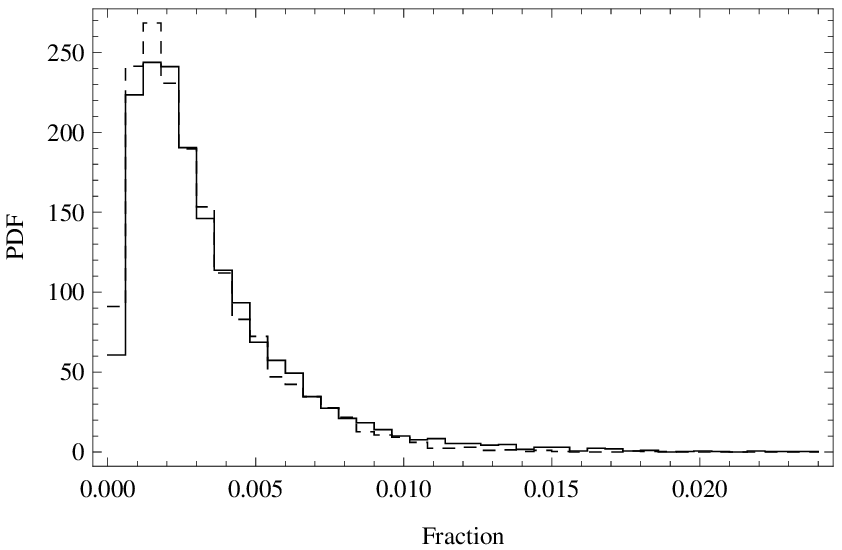}
\includegraphics[width=3.4in]{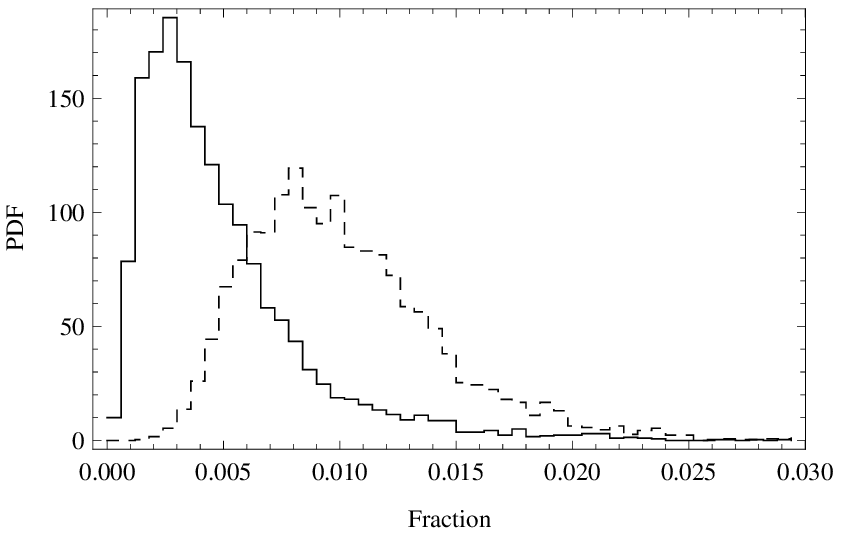}
\caption{\label{Fig:PDFfrac}The PDFs of the gas fraction in halos 
above the minimum cooling mass (solid line), and the gas fraction in
minihalos, i.e., halos below the minimum cooling mass (dashed line) at
$z = 20$ for the case with the $\vbc$ effect (upper panel) and without
the effect (lower panel).}
\end{figure}

To better understand global properties of the first objects we
calculate probability distribution functions of the total gas fraction
in halos as well as the gas fraction in halos above the minimum
cooling mass $M_{\rm min}$. As noted earlier, these gas fractions are
affected by the distribution of relative velocities as well as the
distribution of large scale overdensities $\bar{\delta}_R$. We
consider the PDF of gas fractions inside spherical regions (``cells'')
of radius $R = 3$ Mpc, which are small enough that $\vbc$ can be
treated as roughly constant over a cell. We obtain the PDFs by running
a Monte Carlo simulation that generates random values of $\vbc$ using
Eq.~(\ref{vbcPDF}) and of the large-scale overdensity within the cell
$\bar\delta_R$ using a Gaussian of variance $S'(R)$. In
Fig.~\ref{Fig:PDFfrac} we show the PDFs of the gas fractions in halos
above and below the minimum cooling mass $M_{\rm min} \approx 6\times
10^5 M_{\odot}$ at $z=20$. We also show the same distributions for the
case with no $\vbc$ effect. The figure shows how minihalos would be
dominant at $z=20$ (by a factor of 2 compared to galaxies), but since
$\vbc$ has a larger effect on the minihalos, it makes the gas content
roughly equal between galaxies and minihalos at this redshift. Each
PDF has a non-Gaussian extension towards high fractions (in fact, the
distribution is approximately lognormal). Thus, the peak of the PDF is
significantly lower than the mean value; without the relative velocity
it is 0.002 for galaxies and 0.008 for minihalos, and $\vbc$ moves it
to $\sim 0.0015$ for both. Also, the relative velocities reduce the
full width at half maximum from 0.004 (galaxies) and 0.008 (minihalos)
to 0.003 for both.

\section{Conclusions}
\label{sec6}

We have shown that the relative velocity of baryons and dark matter
has a significant impact on the properties of the first bound objects
and has to be considered in detailed studies of the epoch of
reionization and especially earlier epochs. The supersonic motion of
the baryonic fluid relative to the underlying potential wells created
by the dark matter causes advection of small-scale perturbations by
large-scale velocity flows, leading to a significant suppression of
gas accretion during halo formation and dramatically increasing the
characteristic mass of gas-rich objects at high redshifts ($z >
10$). In particular, instead of this characteristic filtering mass
being close to the Jeans mass of $2\times 10^4 M_{\odot}$ at $z=20$,
it varies among various regions from this value up to $\sim 10^6
M_{\odot}$, with a $1\sigma$ value (and global average) around $M_F =
2\times 10^5 M_{\odot}$, i.e., an order of magnitude higher than
without the relative velocity effect.

The relative velocity effect also modifies the star formation history,
delaying star formation and causing significant spatial
fluctuations. However, since the minimum mass for H$_2$ cooling
($\approx 6\times 10^5 M_{\odot}$ at $z=20$) is somewhat higher than
the average $M_F$, the suppression effect of $\vbc$ is limited to
about a factor of 1.6 at $z=20$ (added on top of the spatially-uniform
factor of 1.2 from the still-depressed baryon perturbations on large
scales), compared to a much larger effect (a factor of 3.3) on the gas
fraction in star-less gas minihalos. The importance of the relative
velocity grows steadily with redshift, so that at $z=40$ the
suppression factors due to $\vbc$ increase to 2.5 for galaxies (on top
of a pre-existing factor of 1.5) and 21 for minihalos.

In our detailed treatment, we included the spatial variation of the
baryonic sound speed, the suppression of baryonic perturbations on
large scales, and the effect of the relative velocity, through the
modified power spectrum, both on the halo mass function and the
internal gas fractions in halos. In order to gauge the induced spatial
variability, we further calculated the full probability distribution
functions of the characteristic mass and of gas fractions inside of
the first collapsed halos. These results are important for
understanding of the relative velocity effect on large scales, and we
plan to study them further.

Our results significantly extend the work done recently by
\citet{Dalal10}. For example, we find a suppression factor due to $\vbc$ 
at $z=20$ of 1.6 and 3.3, for star-forming halos and minihalos,
respectively. In their approach \citet{Dalal10} did not separate these two categories,
and found a factor of 2.5 suppression in the collapsed fraction, which
under their approximation can be interpreted as a suppression of star formation. In our work we removed this and many other approximations used in \citet{Dalal10}. Comparing to our
work, we expect that their calculation of Lyman-$\alpha$ flux fluctuations is qualitatively correct but may be somewhat overestimated and requires a more detailed analysis.

As we were finishing this paper, two simulation papers appeared on the
effect of $\vbc$ at high redshift \citep{Maio,Stacy}. While both found
a small suppression of star formation, their results appear at first glance to show a
smaller suppression effect than we predict. This difference is not surprising if we note that these simulation papers
focused on star-forming halos around $z\approx 15$, while the largest
effects that we find occur for star-less minihalos at higher
redshifts. At $z>20$, \citet{Maio} find tens of percents difference in the gas fractions,
although the statistical errors are large. \citet{Stacy} find a delay in gas collapse by $\Delta a/a=0.14$ for $v_{\rm bc}/\sigma_{\rm vbc}=1$.
We also note that the choice of initial conditions should be carefully considered: standard initial condition codes do not properly treat the separate baryonic and dark matter perturbations or the gas temperature perturbations, leading to a filtering mass that is too high by a factor of $\sim 2$ at $v_{\rm bc}=0$ \citep{NYB}; as such they may underestimate the effect of relative velocities.

The simulations by \citet{Maio} and \citet{Stacy} clearly represent a very important step and will serve as a good foundation for simulations with larger boxes and improved initial conditions. Eventually we hope that simulations including $v_{\rm bc}$ will advance to the point where improved fitting functions for the local halo mass function and gas mass fraction become available.

We note that various feedback effects may reduce or mask some of the
effect of the relative velocity. For galaxies, local feedback from
star formation may effectively raise the minimum halo mass for star
formation (except for the very first generation of stars). The
possibilities include supernova feedback as well as radiative feedback
acting via photoheating and photoevaporation or suppression of
H$_2$ formation, although ``positive'' feedback due to X-ray ionization enhancing H$_2$ formation has also been suggested \citep{hrl96, hrl97}. For minihalos, astrophysical heating,
e.g., from an early X-ray background, may heat the gas and raise the
filtering mass above the value due to $\vbc$. There are many unknowns,
but these various effects could begin to be significant by $z \sim
20$, and very likely by the time of significant cosmic reionization. Still, the relative velocity between baryons and dark matter is the
main determinant of the gas content of halos at the highest redshifts.

\section*{Acknowledgments}
D.T. and C.H. are supported by the U.S. Department of Energy
(DE-FG03-92-ER40701) and the National Science Foundation
(AST-0807337). C.H. is supported by the David and Lucile Packard Foundation.
R.B. is supported by Israel Science Foundation grant 823/09.


\label{lastpage}

\end{document}